\documentclass[preprint]{aastex}
\usepackage{natbib}
\usepackage{subfigure}
\title{A Method for Data-Driven Simulations of \\Evolving Solar Active Regions}

\author{Mark C. M. Cheung \& Marc L. DeRosa \affil{Lockheed Martin Solar and Astrophysics Laboratory \\3251 Hanover St. Bldg/252, Palo Alto, CA 94304, USA.}}
\begin{abstract}
We present a method for performing data-driven simulations of solar active region formation and evolution. The approach is based on magnetofriction, which evolves the induction equation assuming the plasma velocity is proportional to the Lorentz force. The simulations of active region coronal field are driven by temporal sequences of photospheric magnetograms from the Helioseismic Magnetic Imager (HMI) instrument onboard the Solar Dynamics Observatory (SDO). Under certain conditions, the data-driven simulations produce flux ropes that are ejected from the modeled active region due to loss of equilibrium. Following the ejection of flux ropes, we find an enhancement of the photospheric horizontal field near the polarity inversion line. We also present a method for the synthesis of mock coronal images based on a proxy emissivity calculated from the current density distribution in the model. This method yields mock coronal images that are somewhat reminiscent of images of active regions taken by instruments such as SDO's Atmospheric Imaging Assembly (AIA) at extreme ultraviolet wavelengths.
\end{abstract}

\begin{document}
\section{Introduction}
Understanding the structure and evolution of the solar coronal magnetic field
is an important component in understanding space weather.  These dynamics are
a consequence of the transport of energy through the photosphere and into the
chromosphere and corona, and of the subsequent reorganization by the coronal
magnetic field in response to this energy input.  Some of this energy is used
to accelerate the solar wind, some is used to heat plasma trapped in closed
coronal loops (after which it is emitted as radiation), and some is used to
power flares and eruptive phenomena as coronal mass ejections (CMEs).

Coronae above active regions have received particular attention~\citep[e.g.,][]{RegnierPriest:2007, Kazachenko:2012} because these are the sites that
produce the most numerous and strongest flares, and are most often associated
with CMEs and particle acceleration events.  Additionally, active region
coronae are copious emitters in the extreme ultraviolet (EUV) and x-ray
wavelengths of light, with emissions typically taking the form of loop-shaped
structures that are assumed to trace out the trajectories of coronal magnetic
field lines.

Despite considerable progress in observing the structure and evolution of the
solar corona, the root causes of many phenomena remain elusive.  Much of this
uncertainty stems from the fact that it is difficult to accurately map out the
three-dimensional geometry of the coronal magnetic field, and to observe its
temporal evolution.  Direct measurements of the coronal field are sometimes
available off the solar limb~\citep[e.g.,][]{Lin:CoronalFieldMeasurements,BrosiusWhite:2006, Tomczyk:CoronalPolarization,Liu:CoronalIR}, but these measurements often lack
sufficient spatial or temporal resolution, and suffer from line-of-sight
confusion, thereby making proper interpretation somewhat challenging.

As a result of these difficulties, much effort has been put toward modeling
the coronal magnetic field using photospheric magnetograms and/or coronal loop
imagery as constraints.  Many models make use of photospheric measurements of
the magnetic field as boundary conditions.  These models include potential
(current-free) field extrapolations which provide a general idea of the global
connectivity, force-free models which accommodate currents but assume the
corona is in static force balance, and more physically realistic
magnetohydrodynamic (MHD) models which capture many more of the important
dynamical processes.

Of these various coronal magnetic field models, only the potential-field
source-surface (PFSS) model can presently be computed in real time in a
forecasting (predictive) capacity. However, because the active region coronae
of particular interest are those that likely contain significant currents, the
applicability of the PFSS model for understanding energy buildup and release in active regions
is limited, restricting the PFSS model to instances where only visualizing the
large-scale (global) geometry of the coronal field, or providing contextual
magnetic fields surrounding active regions during relatively quiescent periods~\citep{Riley:2006}, is needed.

At the other end of the spectrum of coronal field models are MHD simulations
~\citep[e.g.,][]{Mikic:1999}. These simulations are the most sophisticated in
terms of physical realism, but, even with today's supercomputing technology,
remain computationally expensive for active-region-sized domains at
appreciable resolution.  Additionally, because not all of the necessary
photospheric boundary conditions (e.g., gas pressures and electric fields) are
measured directly, and because they still require parameterized models to
bypass treatment of the microphysics (e.g., coronal heating functions) which
are not always well known, even MHD models are not free from assumptions.

In between the PFSS and MHD models lie the class of nonlinear force-free field
(NLFFF) models of the corona. NLFFF models contain no dynamics, but do allow
for field-aligned electric currents and thus enable measurements of magnetic
free energies and magnetic helicity. Fast algorithms that extrapolate the
magnetic field upward into the corona have been developed, though with mixed
results in terms of their ability to reliably reproduce the observed coronal
field structures~\citep{Schrijver:NLFFF,DeRosa:NLFFF}.  Yet, because
the electric current systems that occupy the corona tend to be isolated in and
around active regions, NLFFF models seem particularly well suited for detailed
studies of the build-up and release of energy within the corona.

One NLFFF solution technique, the magnetofrictional (hereafter MF) scheme~\citep{YangSturrockAntiochos:Magnetofriction,CraigSneyd:Magnetofriction}, involves integrating a model forward in such a
way as to reduce the magnetic stress in the model.  This is achieved
by assuming that the inductive velocity is parallel to the Lorentz force,
which over time causes the magnetic field in the model to relax to a
force-free state.  This method has previously been employed as a way to create
a NLFFF from a non-force-free initialization, either via the specification of
the vector field on the lower boundary of the domain~\citep[][]{Valori:NLFFFExtrapolation,Valori:NLFFFExtrapolationII} or via the insertion of a
flux-rope structure into an otherwise potential field~\citep[e.g.,][]{vanBallegooijen:2004,Bobra:FluxRopeInsertion, Savcheva:2009,Savcheva:2012}.

%% please double-check that the info mentioned in the next paragraph is
%% reasonably accurate

Alternatively, one may drive a MF model by allowing the lower-boundary data to
change with time.  Long-term studies of the formation and subsequent evolution
of filament channels using MF schemes have been performed by~\citet*{vanBallegooijenPriestMackay:MeanFieldModel} and~\citet*{Mackay:FilamentChirality},
in which it was found that potential-field models did not result in sheared
arcades associated with filament channels.  Instead, evolving a MF model over
multiple rotations was needed to recover the skew of coronal fields observed
above the polarity inversion lines along which filaments appear.  Agreement
between observations of filament handedness and their modeled counterparts was
found to improve as the model was run for longer (multiple months) periods of
time~\citep*{Yeates:2007,Yeates:2008,Yeates:2009a,Yeates:2010}, indicating that the build-up and transport of
energy over these time scales is likely an important factor.  The same
evolving MF scheme was used in studies investigating the processes and
frequencies by which filaments lift off~\citep{MackayVanBallegooijen:2006a,
MackayVanBallegooijen:2006b,Yeates:2009b,Su:2011}, the results of which showing
that the changing magnetic geometry associated with neighboring bipolar active
regions is critical in determining how the upward ejection of flux ropes
occurs.

Until now, detailed NLFFF models of active region coronae have primarily been
constructed only from single, isolated vector magnetograms, and therefore
incorporate no information about the prior evolution of the photospheric and
coronal magnetic field.  With the advent of the Helioseismic and Magnetic
Imager~\citep[HMI;][]{Scherrer:HMI,Schou:HMI} on board NASA's Solar
Dynamics Observatory (SDO), time series of photospheric vector magnetogram
data are now available at relatively high resolution ($0.5"$ pixels) and
cadence (every 12 min), with unprecedented spatial and temporal coverage.
Such data enable detailed models of active regions to be constructed, wherein
the MF scheme is utilized to advance the models in time in response to driving
from time series of photospheric magnetograms.  This technique provides a way
to accommodate the (possibly widespread) instances where the prior history of
the state of the coronal magnetic field factors into the determination of its
future state.

We show in a series of articles that it is feasible to use a MF
approach with a time-evolving lower boundary condition to model the coronal evolution of an active
region over a week-long period of time, and that ejections of magnetic flux can be
driven in a model corona using this approach.  In this article, the first in
the series, we outline the numerical methods and demonstrate the scheme on
NOAA Active Region (AR) 11158 using idealized boundary conditions.  AR 11158,
which was on disk in mid Feb.~2011 and produced several major flares, is
followed for a significant fraction of its disk passage.

\section{Method}
Our approach is based on the~\emph{magnetofriction method}, which was introduced by~\citet{YangSturrockAntiochos:Magnetofriction} and~\citet{CraigSneyd:Magnetofriction} to examine the relaxation of magnetic configurations toward force-free states. All MF models
share a common assumption, namely
\begin{equation}
\vec{v} = \frac{1}{\nu}\vec{j}\times\vec{B}.\label{eqn:magfriction}
\end{equation}
\noindent In other words, the plasma velocity $\vec{v}$ is everywhere proportional to the Lorentz force $\vec{F} = (4\pi)^{-1}\vec{j}\times\vec{B}$, where the current density is defined here as $\vec{j}=\nabla\times\vec{B}$, and $\nu$ is a specified magnetofrictional coefficient. Given this chosen form for the velocity field, the magnetic field $\vec{B}$ is evolved in accordance with the magnetohydrodynamic induction equation
\begin{equation}
\frac{\partial \vec{B}}{\partial t} = \nabla \times (\vec{v\times}\vec{B} - \eta\vec{j} ),\label{eqn:induction}
\end{equation}
\noindent where $\eta$ is the magnetic resistivity. 

At first glance, the assumption represented by Eq. (\ref{eqn:magfriction}) may appear {\it ad hoc} and somewhat unphysical. However, there are a number of properties which make this assumption desirable for modeling the evolution of magnetic fields in a magnetically dominated (low plasma $\beta$) regime. First of all, the choice of a velocity parallel to the Lorentz force means the magnetic energy in the volume of interest evolves according to
\begin{equation}
\frac{\partial}{\partial t}\left( \frac{\vec{B}^2}{8\pi}\right) + \nabla \cdot \vec{S} = -\left( \frac{1}{\nu}(\vec{j}\times\vec{B})^2 + \eta\vec{j}^2\right),
\end{equation}
\noindent where $\vec{S} = -(\vec{v}\times\vec{B} -\eta\vec{j})\times\vec{B}$ is the Poynting flux of magnetic energy. Since the right-hand side of this equation is never positive (both $\nu$ and $\eta > 0$), it is easy to see that, in the absence of a net Poynting flux (i.e. no energy injection), the total magnetic energy ($M = \int_V B^2/8\pi dV$) inside the volume must decrease monotonically or stay constant in time.

By itself, the property of $\dot{M} \le 0$ is desirable but not necessarily meaningful for studying the temporal evolution of coronal magnetic fields. Another property of the MF method is that, by virtue of the fact that the induction equation is used to advance $\vec{B}$ in time, the topology of the field is preserved (except in the case of high magnetic diffusion). In order words, under ideal MF evolution, important topological quantities such as relative magnetic helicity~\citep{Berger:MagneticHelicity,FinnAntonsen:RelativeHelicity} remain conserved. 
In contrast, this property does not necessarily hold for successive magnetic field configurations generated from NLFFF extrapolation schemes such as the Grad-Rubin~\citep[e.g.][]{Wheatland:FastCurrentFieldIteration} and the optimization~\citep[e.g.][]{Wheatland:OptimizationCode,Wiegelmann:OptimizationCode} methods. Force-free field extrapolations based on these two approaches treat the construction of model fields at two times $t_1$ and $t_2$ as completely independent processes. The application of MF to the construction of force-field fields by~\citet{Valori:NLFFFExtrapolation,Valori:NLFFFExtrapolationII} is similar in that the force-free field obtained at time $t_1$ does not influence the solution at time $t_2$. In this regard, our application of MF to modeling the time evolution of coronal fields should be clearly distinguished from previous efforts that apply MF for force-free field extrapolation. 

One way to physically interpret the MF scheme is to think of Eq. (\ref{eqn:magfriction}) as a simplified momentum equation which specifies the velocity without regard to inertial effects. Another way to interpret the MF scheme is as having a non-linear diffusion term for the magnetic field. We note that magnetic field evolution by MF has the same mathematical character as ambipolar diffusion. Both effects contribute to an electric field in the induction equation proportional to $\vec{B}\times(\vec{j}\times{B})$. As reported by~\citet{BrandenburgZweibel:AmbipolarDiffusion}, ambipolar diffusion leads to the formation of sharp current sheets~\citep[see also][]{CheungCameron:HallMHD}. What is important to note, however, is that both ambipolar diffusion and magnetofriction do not permit reconnection (their electric fields are perpendicular to $\vec{B}$). In order for the topology to change, Ohmic diffusion (which is present in our model) needs to operate in locations of current layers.

\subsection{Implementation}
Our implementation of the MF scheme follows closely the implementation by~\citet{vanBallegooijenPriestMackay:MeanFieldModel}, who used the method to study the formation of filament channels. The major difference between the two methods is in the implementation of the bottom boundary condition, which will be discussed in section~\ref{sec:bottomboundary}.

As an alternative to solving for the magnetic field $\vec{B}$ in time, the method solves the induction equation for the vector potential $\vec{A}$,
\begin{equation}
\frac{\partial \vec{A}}{\partial t} = \vec{v}\times\vec{B} - \eta \vec{j},\label{eqn:evolveA}
\end{equation}
\noindent where by definition $\vec{B} = \nabla \times \vec{A}$. We use a staggered mesh such that Cartesian components of the vector fields $\vec{A}$, $\vec{B}$ and $\vec{j}$ (here defined as $\vec{j}=\nabla\times\vec{B}$)  are defined in the following way: $A_x$ and $j_x$ are defined at the midpoints of cell edges parallel to $\hat{x}$ and $B_x$ is defined at the centers of cell faces with normal vectors parallel to $\hat{x}$~\citep{Yee:StaggeredGrid}. The same arrangement applies to the other Cartesian directions. Numerical values of all components of $\vec{v}$ are defined at the cell vertices. 

Let superscripts $i\in[0,N_x]$, $j\in[0,N_y]$ and $k\in[0,N_z]$ denote the coordinate locations of cell vertices in the $x$, $y$ and $z$ directions, respectively. The length of the cell edges in the three directions are $\Delta x$, $\Delta y$ and $\Delta z$. In order to evaluate $\vec{B}=\nabla\times\vec{A}$, second-order spatial derivatives of $\vec{A}$ of the form
\begin{eqnarray}
B_x^{i,j+\frac{1}{2},k+\frac{1}{2}} &=& \frac{A_z^{i,j+1,k+\frac{1}{2}} - A_z^{i,j,k+\frac{1}{2}}}{\Delta y} \\
                  &-& \frac{A_y^{i,j+\frac{1}{2},k+1} - A_y^{i,j+\frac{1}{2},k}}{\Delta z}
\end{eqnarray}
\noindent are used. Evaluation of $\vec{j}$ also employs second-order spatial derivatives:
\begin{eqnarray}
j_x^{i+\frac{1}{2},j,k} &=& \frac{B_z^{i+\frac{1}{2},j+\frac{1}{2},k} - B_z^{i+\frac{1}{2},j-\frac{1}{2},k}}{\Delta y} \\
              &-& \frac{B_y^{i+\frac{1}{2},j,k+\frac{1}{2}} - B_y^{i+\frac{1}{2},j,k-\frac{1}{2}}}{\Delta z}.
\end{eqnarray}
\noindent Derivatives for the $y$ and $z$ directions use the same scheme. To evaluate the magnetofrictional velocity $v$ at cell vertices,
linear interpolation of $\vec{j}$ and $\vec{B}$ is used. Furthermore, the same interpolation scheme is used to evaluate the $\vec{v}\times\vec{B}$ term in Eq.(\ref{eqn:evolveA}).

For the simulations presented here, the functional form of the frictional coefficient is given by
\begin{equation}
\nu = \nu_0 B^2(1- e^{-z/L})^{-1},\label{eqn:nu}
\end{equation}
\noindent where $\nu_0 = 10$ s Mm$^{-2}$, $L = 10$ Mm and $z$ is the height above the bottom boundary (i.e. photosphere). As pointed out by~\citet{Low:CraigSneydCriticism}, some past models that use the magnetofriction approach have adopted a constant magnetofrictional coefficient $\nu$~\citep[e.g.][]{CraigSneyd:ParkerProblem} while assuming a line-tied, rigid bottom boundary where flows vanish. This could lead to an undesirable mismatch between the boundary condition and the magnetofrictional velocities that in turn result in unphysical current sheets near the bottom boundary. The height dependent profile of $\nu$ in Eq.(\ref{eqn:nu}) is chosen such that magnetofrictional velocities smoothly transition to zero towards $z=0$.

It is worthwhile to estimate the relevant timescales based on the above choice of the magnetofrictional coefficient. Since the MF induction equation is nonlinear, it is difficult to estimate relaxation timescales with general validity. To estimate characteristic timescales, we distinguish between the local adjustment timescale $\tau_l$ and the disturbance propagation timescale $\tau_p$. From Eqs. (\ref{eqn:magfriction}) and (\ref{eqn:induction}), we define the local adjustment timescale by comparing the order of magnitude of the terms in the induction equation. Assuming $|\frac{\partial \vec{B}}{\partial t}| \sim B/\tau_l$ and $|\nabla\times\vec{B}| \sim B/l$, the local adjustment timescale is 
\begin{equation}
\tau_l = \nu_0 l^2(1-e^{-z/L})^{-1},
\end{equation}
\noindent where we have made use of Eq. (\ref{eqn:nu}). In the above equation, $B$ is the local field strength and $l$ is the spatial scale over which $B$ varies.

The disturbance propagation timescale $\tau_p$ measures the time taken for a change in the photosphere (i.e. z=0) to reach a height $z=h$. We define it as
\begin{equation}
\tau_p (h) = \int_0^h \frac{dz}{|\vec{v}|} =  \int_0^h \frac{\nu_0 l}{1-e^{-z/L}} dz,
\end{equation}
\noindent where we have again assumed that $|\nabla \times \vec{B}| \sim B/l$. Estimates of both timescales depend on the choice of $l$, which is typically smaller at lower heights where there is a greater degree of mixed polarity field. In potential field models~\citep[e.g.][]{Gary:Extrapolation}, the amplitude of the Fourier coefficient at spatial wavenumber $k$ scales as $b_k \sim e^{-kz}$. So at a height $z$ in a potential field, one may expect to find field gradients over spatial scales of $l \sim z$. If one adopts $l \sim z$, the local adjustment times at $z=1$ Mm and $z=50$ Mm (typical range of loop heights in an AR) would be $\tau_l \sim 2$ min and $\tau_l \sim 7$ hr, respectively. With the same assumption $l \sim z$, the disturbance propagation times to heights of $z=1$ Mm and $z=50$ Mm are $\tau_p \sim 2$ min and $\tau_p \sim 4$ hr, respectively. However, these numbers for both the local adjustment time and disturbance propagation times should be considered as upper limit estimates (for the corresponding height) since we assumed that at any height, the typical spatial scale of the magnetic field gradient is $ l\sim z$. While this is valid for a potential field, MF evolution can create sharp structures with $l \ll z$. It is worth noting that relaxation timescales depend linearly on $\nu_0$. For smaller (larger) values of $\nu_0$, the model coronal field evolves over shorter (longer) timescales in response to photospheric changes and vice versa. Simulations performed with the present choice of $\nu_0$ allows the model coronal field to accumulate and store free magnetic energy over the course of days while still allowing for flux ropes to be created and ejected over shorter timescales (see section~\ref{sec:results}).

The magnetic diffusivity $\eta$ in Eq. (\ref{eqn:evolveA}) has contributions from three different terms. The first contribution is a constant, spatially uniform resistivity of $\eta_0 = 200$ km$^2$ s$^{-1}$. The second contribution has a functional form that is designed to facilitate diffusion in regions of high current density. It is given by
\begin{equation}
\eta_1 = \eta_0 \frac{10 \zeta}{1 + \exp\{-(\zeta-0.1)/0.01\}},
\end{equation}
\noindent where $\zeta = j^2 B^{-2}\Delta^2$ and $\Delta = \rm{min}\{\Delta x, \Delta y, \Delta z\}$. The third contribution is a hyperdiffusivity-like scheme. It is similar to the hyperdiffusivities used by~\citet{Caunt:Hyperdiffusivities} in the sense that the diffusivity is proportional to the ratio of the third and first derivatives of the quantity being diffused. This type of diffusivity is efficient at suppressing oscillations at the grid scale. The form of the hyperdiffusivity for $A_x$ is

\begin{equation}
\eta_{{\rm hyp},x}^{i+\frac{1}{2},j,k} = v \Delta |A_x^{3}/A_x^{1} |,
\end{equation}
\noindent where $A_x^{3} =  j_x^{i+3/2,j,k} - 2 j_x^{i+\frac{1}{2},j,k} + j_x^{i-\frac{1}{2},j,k}$, $A_x^{1} = j_x^{i+\frac{1}{2},j,k}$ and 
$v  = {\rm max}\{|\vec{v}|^{i,j,k}, |\vec{v}|^{i+1,j,k} \} $. The same scheme is used for hyperdiffusivities in the remaining Cartesian directions.

A second-order midpoint method is used to explicitly evolve $\vec{A}$ forward in time, with each update consisting of a half-step and a full-step. The adaptive time step is determined by a CFL-like condition given by
\begin{equation}
\Delta t = \frac{1}{4}{\rm min}\{\Delta/v_{\rm max} , \Delta^2 / \eta_{\rm max}\},
\end{equation}
\noindent where $v_{\rm max}$ and $\eta_{\rm max}$ are the maximum values of the magnetofrictional velocity and total magnetic diffusivity over the computational domain. 
%\begin{eqnarray}
%v_x^{i,j,k} &=& \nu^{-1}[(j_y^{i,j-\frac{1}{2},k} + j_y^{i,j+\frac{1}{2},k})(B_z^{i,j+\frac{1}{2},k+\frac{1}{2}} + B_z^{i+1,j+\frac{1}{2},k+\frac{1}{2}}) - \\
%& & (j_z^{i,j,k-\frac{1}{2}} + j_z^{i,j,k+\frac{1}{2}})(B_y^{i+\frac{1}{2},j,k+\frac{1}{2}}+B_y^{i+\frac{1}{2},j+1,k+\frac{1}{2}}) ]
%\end{eqnarray}

\subsection{Initial condition}
For each simulation, the initial condition for $\vec{A}$ corresponds to a potential field configuration (i.e. current-free). This is generated by means of a potential field extrapolation using the vertical component of the first magnetogram in the sequence. The potential field extrapolation is performed with a Fourier method and assumes periodic boundary conditions in the horizontal directions. Since the surface field in the first frame of $\bar{B}_z$ consists of mostly pre-existing weak field before an active region emerges, the periodicity assumption should not play a big role in determining the dynamical evolution of the AR as it emerges.

\subsection{Boundary conditions}

\subsubsection{Top and side boundaries}
At the top and side boundaries of the Cartesian domain, the magnetofrictional velocity $\vec{v}$ values at the boundary cell vertices are chosen to be the same values as those defined on cell vertices one layer deep in the computational domain. The magnetic field $\vec{B}$ is normal at the boundaries and the transverse component is set to vanish. The normal component of $\vec{j}$ is symmetric across the boundaries and the transverse components are antisymmetric across the boundaries.

\subsubsection{Bottom boundary}
\label{sec:bottomboundary}
A time-dependent boundary condition based on temporal sequences of magnetograms is imposed at the bottom of the computational domain (i.e. photosphere) to drive the evolution of the magnetic field in the model corona ($z> 0$). Mathematically, the choice of a boundary condition in this time-dependent problem involves a choice of the three components of the electric field (i.e. $\vec{E}=-\partial_t \vec{A}$) at $z=0$. From Eq. (\ref{eqn:evolveA}), one sees that this corresponds to a choice of the photospheric (i.e. at $z=0$) distributions of $\vec{v}$, $\vec{B}$, $\vec{j}$. A consistent boundary condition, however, does not allow one to impose arbitrarily all three components of all three vectors. Consider, for example, the horizontal components of the electric field. To specify the boundary condition for these two components, one must specify all three components of $\vec{B}$ and $\vec{v}$ as well as the horizontal components of $\vec{j}$. This choice automatically constrains the remaining component of the electric field (i.e. $-\partial_t A_z$) since  $j_z(z=0) =[ \partial_x B_y - \partial_y B_x]_{z=0}$. This examples demonstrates that an appropriate boundary condition for the problem permits a choice of only two out of three components of the boundary electric field. This property is reflected in the choice of the staggered grid. At each boundary on this grid, only the transverse components of $\vec{A}$ (as well as their time-derivatives) are defined.

Having established that only $E_x$ and $E_y$ need to be specified, one still has to tackle the problem of how these two quantities can be retrieved from observed temporal sequences of magnetograms. This problem was recently analyzed in detail by~\citet{Fisher:EstimatingElectricFields}. An important conclusion from their work is that retrieval of the photospheric electric field from temporal sequences of vector magnetograms taken at one height is an ill-posed problem for which the solution is not unique. Where available, magnetograms at various heights in the atmosphere may help resolve the problem. At present however, there is no instrument capable of providing magnetograms at different heights at the same level of spatial coverage, resolution and time cadence that SDO/HMI can offer. Since SDO/HMI only provides magnetograms at one layer in the photosphere, one has to find ways to cope with the missing information.

The simulations presented here are driven by electric fields derived from time sequences of SDO/HMI \emph{longitudinal} (i.e.line-of-sight) magnetograms under a number of assumptions. Since the purpose of this paper is to introduce the data-driving method, the simulations presented here do not attempt to be fully realistic. In order to drive a simulation in a realistic fashion, photospheric electric fields faithful to the actual electric fields operating on the Sun must be used. The retrieval of such electric fields constrained by temporal sequences of vector magnetograms (e.g. from SDO/HMI) is still a research problem~\citep{Fisher:EstimatingElectricFields,Fisher:EfieldsWithDoppler} and the evaluation of the quality of this inversion process is outside the scope of this paper. The application of electric fields derived from HMI vector magnetograms, and a critical assessment of the realism of the model coronal field is the topic of the second article in this series.

For each active region simulation, we prepare a time sequence of input magnetograms $\bar{B}_z$, extracted from the series of full-disk SDO/HMI magnetograms. This quantity is used to specify the distribution of the vertical magnetic flux density at the bottom boundary of the Cartesian simulation domain throughout the simulated time period. The central position of the $z=0$ plane in the domain is chosen to co-rotate with the Sun at a rate (for fixed latitude) given by the empirical differential rotation profile for surface magnetic features obtained by~\citet{SnodgrassUlrich:DifferentialRotation}. Each magnetogram in the input sequence $\bar{B}_z$ spans $30.7^\circ$ in heliographic latitude and longitude. The models here~\emph{neglect} the effect of curvature over the $30.7^\circ\times 30.7^\circ$ patch. A {\it plate carr\'ee} projection is used for remapping the magnetograms. With $512\times 512$ pixels in each direction, the effective grid spacing at the bottom boundary of the domain is $\Delta x = \Delta y = 728$ km, corresponding to one-arcsecond grid spacing in the plane of the sky (at disk center). One further assumption made is that the surface field on the Sun is purely radial. This assumption is used to convert the longitudinal magnetogram values into values for $\bar{B}_z$. To conserve flux balance in the magnetogram remapping process, the effects of foreshortening of surface area elements away from disk center are taken into account. Consecutive frames of the remapped magnetograms $\bar{B}_z$ are generated at a regular cadence of $4.5$ minutes from SDO/HMI longitudinal magnetograms. This is a lower cadence than the 90 second cadence offered by SDO/HMI but sufficiently high to have a non-negligible impact on the memory and I/O requirements of the MF code. 

Given the sequence of input vertical magnetograms $\bar{B_z}$, its relation to $\vec{E}_h = (E_x,E_y)$ is given by the vertical component of the induction equation

\begin{equation}
\frac{\partial \bar{B}_z(x,y) }{\partial t} = -\hat{z}\cdot ( \nabla \times \vec{E}_h).
%-\left[ \frac{\partial E_y}{\partial x} - \frac{\partial E_x}{\partial y}\right].
\label{eqn:curlE}
\end{equation}
\noindent In order to solve for $\vec{E}_h$, another relation must be specified. For example, as is done here, specifying the horizontal divergence of the electric field suffices:
\begin{equation}
D(x,y) = \nabla_h \cdot \vec{E}_h.\label{eqn:divE}
\end{equation}
\noindent Given a chosen form of $D(x,y)$, Fourier transforms are used to solve Eqs. (\ref{eqn:curlE}) and (\ref{eqn:divE}) for $\vec{E}_h$.

Two simulations with identical initial conditions, driven by the same temporal sequence of photospheric magnetograms $\bar{B}_z$ will nevertheless yield different coronal field configurations if different distributions of $D(x,y)$ are used. Since it is yet unclear how one can unambiguously and robustly retrieve the function $D(x,y)$ (or $\vec{E}_h$) from observations, the simulations in this study should be considered as an exploratory study of coronal field evolution under different assumptions of $D(x,y)$.

% \mcmc{We do not claim that boundary conditions imposed in the simulations are representative of those on the Sun. Instead, the simulations presented should be treated as numerical experiments. In Paper II, we will present results from simulations driven by electric fields retrieved by the PTD method.}

\subsection{Method for synthetic coronal images based on a proxy emissivity}
\label{sec:emissivity}
The MF model does not treat the thermodynamic evolution of plasma in the corona. This means that the MF method does not provide spatial distributions of thermodynamic quantities such as pressure, mass density and temperature throughout the computational domain. Lacking this information, one cannot directly apply atomic physics models~\citep[e.g.~using the CHIANTI package,][]{Dere:CHIANTI} to the synthesis of coronal images at EUV and X-ray wavelengths~\citep[][]{Peter:CoronalHeatingThroughBraiding,Aiouaz:CoronalFunnels,Chen:EITWavesSynthesis,Gudiksen:AbInitioApproach,Mok:ThermalStructureOfSolarARs,WarrenWinebarger:HydroModelingXrayEUV,WarrenWinebarger:StaticDynamicModeling,Lundquist:ARCoronalEmissionPartI,Lundquist:ARCoronalEmissionPartII,Lionello:MultispectralEmission,Hansteen:RedBlueShifts,Zacharias:EjectionOfCoolPlasma,MartinezSykora:ForwardModelingAIA} without making a series of assumptions.

Instead of attempting to synthesize realistic images of the model corona using a first principles approach, we developed a method for calculating proxy emissivities based on the value of the current density-squared ($j^2$) averaged along a magnetic field line. This method allows one to conveniently visualize which field lines are current-carrying.

The method works as follows. Consider a scalar emissivity field $\mathcal{\varepsilon}(x,y,z)$ spanning the computational domain. The `initial state' of the emissivity field is such that $\mathcal{\varepsilon} = 0$ for all points. Now consider a point at the photospheric boundary ($z=0$, i.e. bottom boundary of the computational domain) at position $(x,y,0)$. Using the three-dimensional distribution of $\vec{B}$ at some time $t$, we trace this field line into the computational domain by integrating $dx/B_x = dy/B_y = dz/B_z$. Define for each field-line trajectory the mean-squared current density 
\begin{equation}
\langle j^2 \rangle = L^{-1}\int_0^L j^2 ds,
\end{equation}
\noindent where $L$ is the length of the field line. If a field line crosses one of the side or top boundaries of the computational domain, we set $\langle j^2 \rangle=0$ so that the current in that field line does not contribute to the emissivity. A magnetic field line will typically traverse a number of cell elements in the computational domain. For each of these cell elements, we increment the local value of the emissivity by 
\begin{equation}
d\mathcal{\varepsilon} = G \langle j^2 \rangle\Delta x \Delta y,
\end{equation} 
\noindent where $G$ is a coefficient which can be uniform in space or be a function of $B$. This procedure is then repeated for a collection of magnetic field lines with footpoint positions equally separated by $\Delta x$ and $\Delta y$ in the horizontal directions.

After performing the aforementioned procedure for a large number (typically of order $10^5$ or $10^6$) of field lines, line-of-sight integrations through $\mathcal{\varepsilon}(x,y,z)$ can be performed through any viewing angle to create synthetic images of coronal loops.

%The field line may end at $z=0$ (i.e. both footpoints are anchored) or it may reach one of the side or top boundaries of the computational domain.

\section{Application to NOAA Active Region 11158}
\label{sec:results}
The set of simulations presented here follow the evolution of NOAA AR 11158 over the course
of multiple days from UTC \textsf{2011-02-10T14:00} to~\textsf{2011-02-15T06:00}. AR 11158 is the source region for the GOES X2.2 flare that took place between \textsf{2011-02-15T01:44} and~\textsf{2011-02-15T02:06}. There are a number of observational and theoretical studies of the X flare and its associated coronal mass ejection ~\citep{Schrijver:AR11158,Jiang:RapidSunspotRotation,Sun:AR11158,Wang:ResponseOfPhotosphereAfterXFlare,Gopalswamy:CMEfromAR11158}. 
The evolution of the AR over the days preceding the flare, however, have not received as much scrutiny. 

The following simulations were performed on a Cartesian grid with $\Delta x=\Delta y=\Delta z=728$ km. All begin with the same potential field initial condition (at UTC \textsf{2011-02-10T14:00})  and are driven by the same sequence of magnetograms ($B_z$). However they differ in the choice of $D(x,y)$, which we have assumed to be of the form 
\begin{equation}
D(x,y) = \Omega B_z(x,y) \label{eqn:divE_choice}
\end{equation}
\noindent where $\Omega$ is a parameter which is kept constant and uniform for each individual run. This functional form of $D(x,y)$ is motivated by the fact that the expression for $\nabla_h \cdot \vec{E}_h = \nabla_h \cdot [-\vec{v}\times\vec{B}]_h$ includes the term $\omega_z B_z$, where $\omega_z = \partial_x v_y - \partial_y v_x$ is the vertical component of the fluid vorticity. Thus a non-zero value of $\Omega$ in Eq. (\ref{eqn:divE_choice}) corresponds to imposing spatially uniform vortical motion (i.e. twisting motion) at the photosphere. Table~\ref{table:ar11158} shows the choices of $\Omega$ for the various runs in this parameter study. The reference run AR11158$\Omega0$ corresponds to the case where no twisting motion is applied at the photosphere.

\begin{table}[h]
\centering
\begin{tabular}{|c|c|}
\hline
Run & $\Omega$ \\
\hline
AR11158$\Omega0$ & 0\\
AR11158$\Omega+0.125$ & $1/8$ turn per day\\
AR11158$\Omega+0.25$ & $1/4$ turn per day\\
\hline
\end{tabular}
\caption{Properties of the various simulation runs for models of AR 11158. $\Omega$ is a parameter that controls the amount of uniform twisting introduced by the time evolution of the photospheric field (see Eq. (\ref{eqn:divE_choice}) for definition).}\label{table:ar11158}
\end{table}

\subsection{Active region morphology}
\begin{figure*}
\centering
\includegraphics[width=\textwidth]{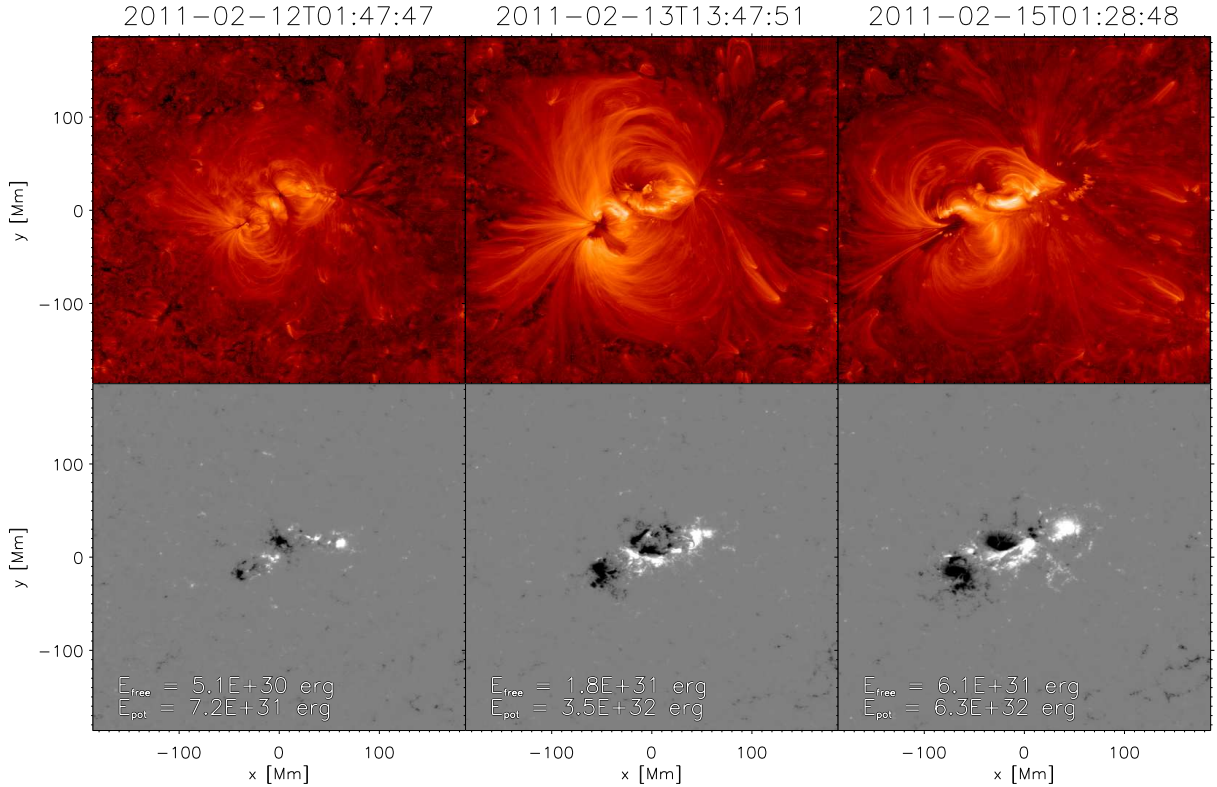}
\caption{Time sequence of synthetic images of coronal loops (upper panels) and underlying photospheric field distributions ($B_z$ scaled between $\pm 800$ Mx cm$^{-2}$, lower panels) of a data-driven simulation of NOAA AR 11158. In this particular simulation, the twisting parameter $\Omega=0$. The synthetic coronal images (shown in logarithmic scaling) were calculated as line-of-sight integrals of the proxy emissivity prescription as described in section \ref{sec:emissivity}.}\label{fig:fig1}
\end{figure*}
\begin{figure*}
\centering
\includegraphics[width=\textwidth]{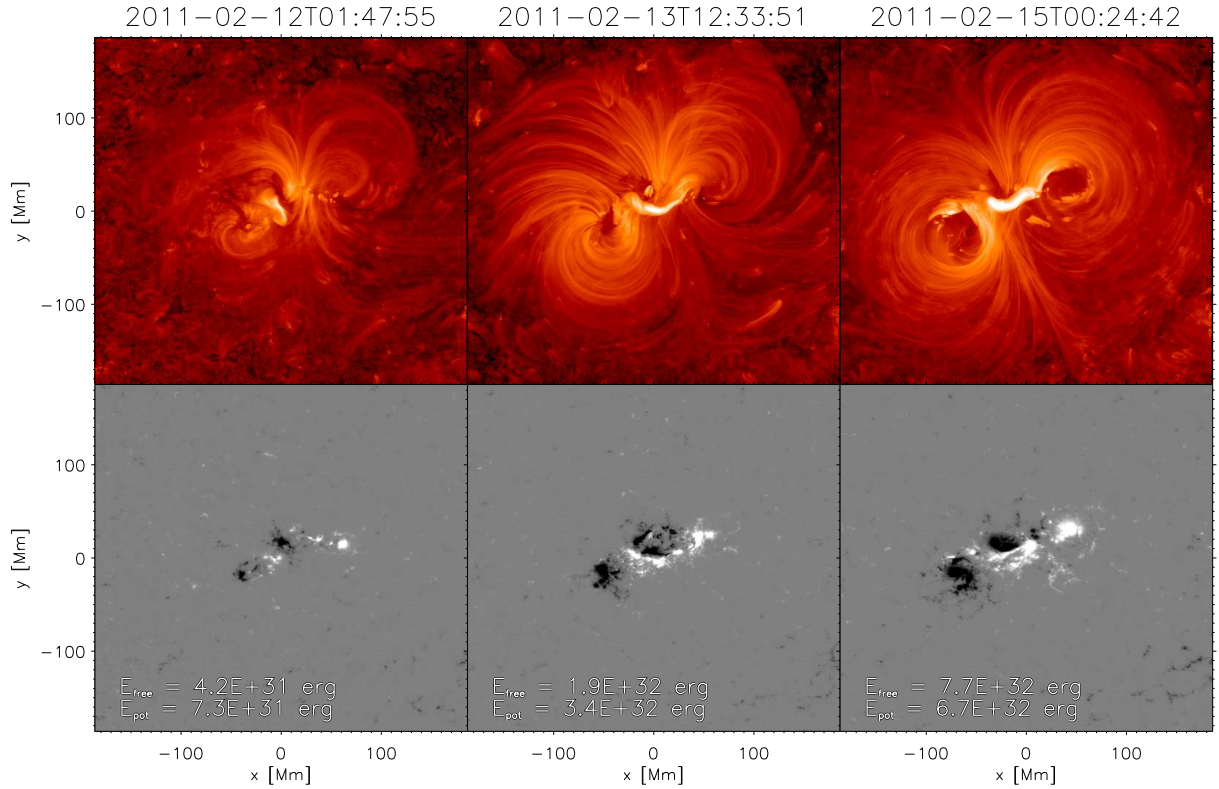}
\caption{Same as Figure~\ref{fig:fig1} but for the run with twisting parameter $\Omega=1/4$ turns per day. The As opposed to the untwisted case, synthetic images for this case show a sheared core (bright structure in the center of the field of view) close the to polarity inversion line between opposite polarities.}\label{fig:fig1_omega}
\end{figure*}

Figure \ref{fig:fig1} shows a time sequence of synthetic coronal images and photospheric magnetograms ($B_z$) from simulation run AR11158$\Omega0$. The former were calculated using the method described in section \ref{sec:emissivity} using $1024\times 1024$ field lines (i.e. four field lines per grid cell at the bottom boundary). The scalar field $\mathcal{\varepsilon}(x,y,z)$ representing a proxy emissivity was calculated assuming $ G = {\rm constant}$ (the actual value of this constant is not important for examining field morphology as long as the same constant is used for different snapshots, as is done throughout this article). The synthetic coronal images shown in the figure represent line-of-sight integrals along vertically-directed rays.

The sequence of images in Fig.~\ref{fig:fig1} gives a sense of how the model coronal magnetic field evolves in response to photospheric driving. At a relatively early stage of the AR emergence (left panel), the magnetic field lines (identified as bright loops in the synthetic coronal images) emanating from the AR are relatively confined. As flux continues to emerge and the AR grows, the area of the image covered by magnetic loops associated with the AR increases as more fieldlines become significantly current-carrying.

Figure~\ref{fig:fig1_omega} shows the sequence of synthetic coronal images and corresponding photospheric magnetograms for the simulation run AR11158$\Omega+0.25$. In this run, a twisting rate of $\Omega=1/4$ turns per day was imposed. Due to this imposed twisting, the field above the sharp polarity inversion line in the central portion of the AR is significantly sheared and appears as a bright core (as a proxy of the current content).

The synthetic coronal images reveal that not all of the magnetic flux within the AR is internally connected. In snapshots for all simulation runs, one finds field lines that emanate from polarity patches within the AR that have conjugate footpoints in pre-existing, diffuse flux patches external to the AR. In fact, as the model AR grows, one finds sets of loops which seem to crawl along the magnetic carpet of the ambient photosphere.

\subsection{Buildup of free magnetic energy}

\begin{figure}
\centering
\includegraphics[width=0.48\textwidth]{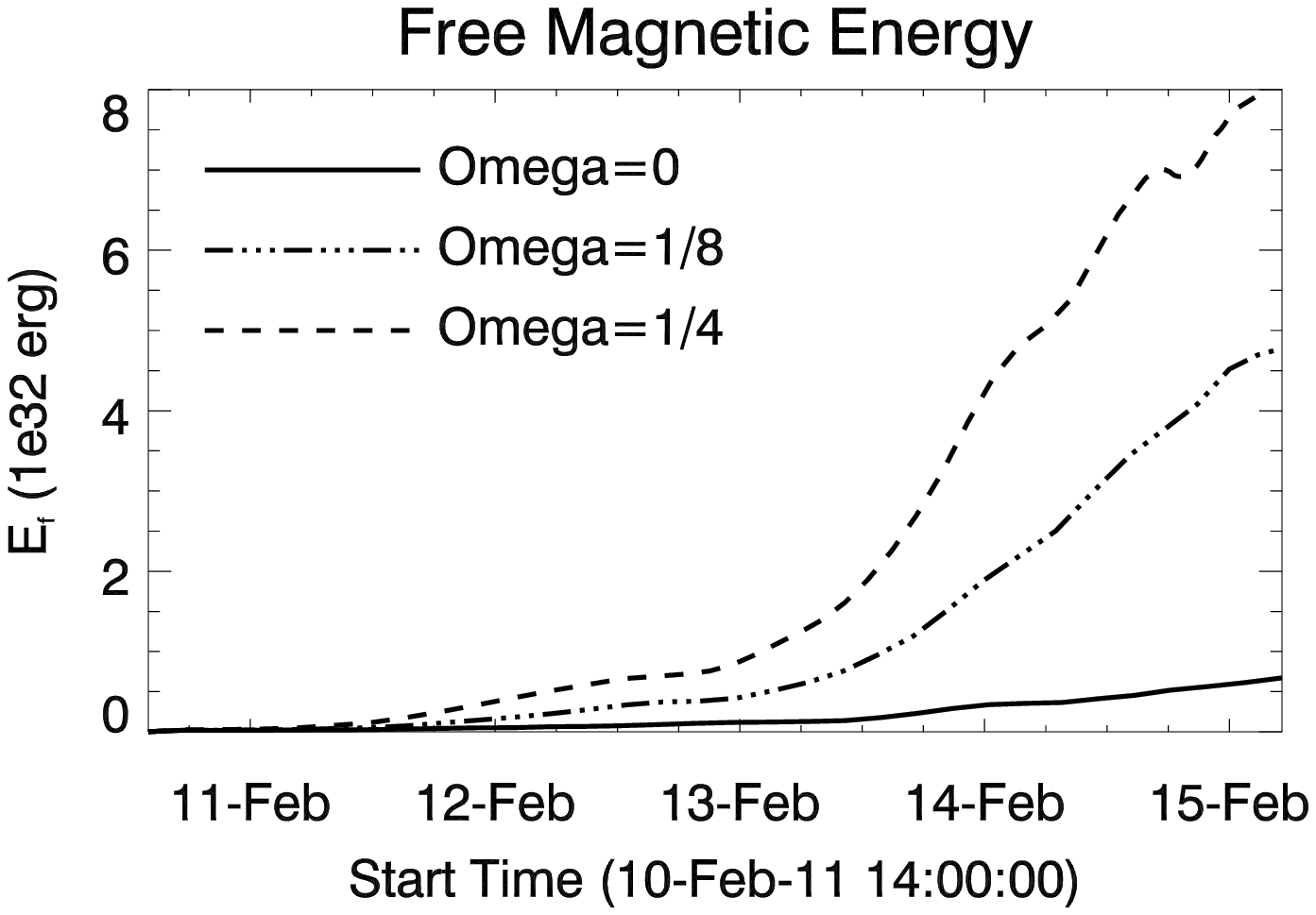}
\caption{Free magnetic energy for various simulation runs. Without imposed twisting ($\Omega=0$, solid line), the free magnetic energy is limited to $\sim 7\times10^{31}$ erg. With imposed twisting ($\Omega = 1/8$ turns per day and $\Omega=1/4$ turns per day, dashed-dotted line and dashed line, respectively), the free magnetic energy is up to an order of magnitude larger. }\label{fig:fig5}
\end{figure}

In the simulation run without imposed twisting (i.e. $\Omega = 0$), the absence of systematic shearing of the AR magnetic fields results in a relatively `quiescent' evolution of the modeled AR. This is reflected in the time plot of the free magnetic energy as shown in Fig~\ref{fig:fig5}. The free magnetic energy is defined as $E_f = E(\vec{B}) - E(\vec{P})$, where $E(\vec{B})$ is the energy of the magnetic field in the magnetofrictional model and $E(\vec{P})$ the energy of the corresponding potential field configuration (specified by the normal field through all boundaries at the same time). For the simulation run with $\Omega=0$, the free energy of the system for the duration of the simulation run is less than $7\times 10^{31}$ erg $\approx 0.07 E(\vec{P})$. In the run with a constant, uniform twisting rate of $\Omega = 1/4$ turns per day, $E_f$ reaches values of $\sim 8\times 10^{32}$ erg $\approx 1.2E(\vec{P})$. So that latter run has a much larger reservoir of free magnetic energy to drive ejections. This is done by means of the creation of a sheared, current-carrying magnetic arcade above the sharp magnetic polarity inversion line in the core of the modeled AR11158. Even though a series of flux rope ejections (see next section) result from the shearing field in the runs with imposed twisting, the free magnetic energy continues to increase. This is simply a result of imposing a twisting rate that is (unrealistically) spatially and temporally uniform, irrespective of the evolutionary phase of the model AR. As reported by~\citet{Jiang:RapidSunspotRotation}, the rotation patterns of the sunspots in this AR are far from uniform. 

\subsection{Recurrent flux rope ejections from sheared fields}

\begin{figure*}
\centering
\includegraphics[width=\textwidth]{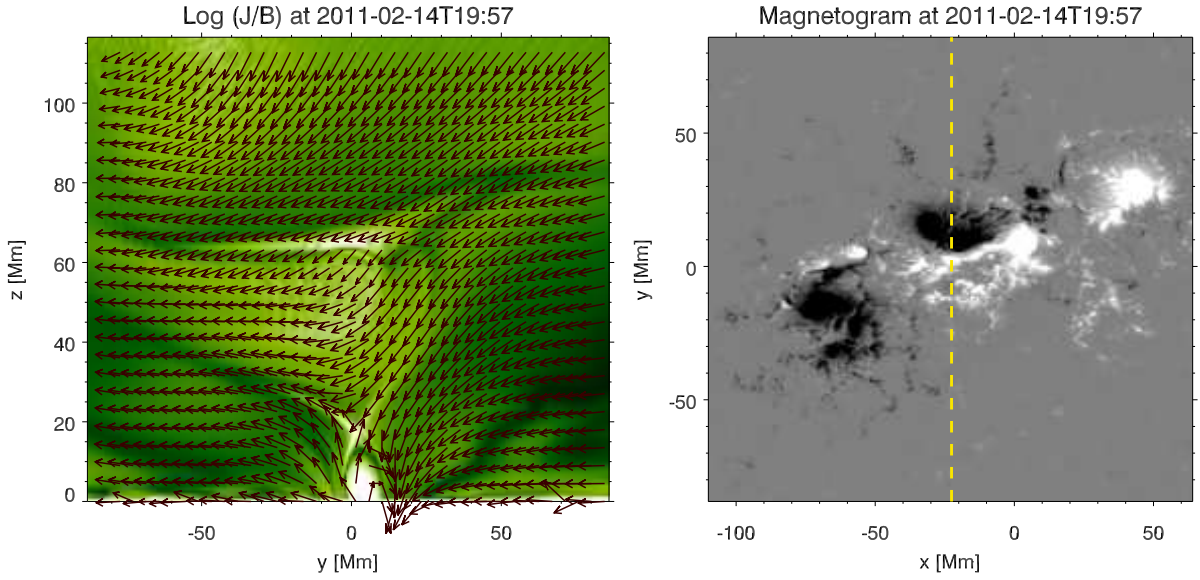}
\caption{The imposed twisting rate of $\Omega = 1/4$ turns per day in simulation run AR11158$\Omega+0.25$ leads to a sheared, current-carrying arcade which is the source region for recurring flux rope ejections. Left panel: Image of vertical cross-section of the normalized current distribution ($|j|/|B|$) in a vertical plane ($x=-22.6$ Mm) above a central portion of the simulated AR in run AR11158$\Omega+0.25$. The vector field shows the transverse component of $\vec{B}$ (normalized by field strength to accentuate the field orientation). Right panel: Corresponding magnetogram of the vertical field at $z=0$. [An animated version of this figure is available.]}\label{fig:fig4}
\end{figure*}

As shown in the left panel of Fig.~\ref{fig:fig4}, flux ropes are formed when magnetic reconnection pinches the upper portion of the sheared arcade. The loss of equilibrium leads to the ejection of the flux rope. Although magnetofriction evolves the magnetic field in a rather simplified manner, such models are still able to capture a number of qualitative aspects of MHD models of coronal field evolution. For instance, the creation of flux ropes from a sheared arcade and their subsequent ejection due to loss of force equilibrium is present MHD simulations with shear~\citep{MikicLinker:ShearedArcades} and converging flows~\citep{Amari:CMEPartI} about the polarity inversion line of the model bipole. For the particular case of the X2.2 flare and the associated eruption of AR 11158, \citet{Schrijver:AR11158} used the MHD simulations of~\citet{Aulanier:FormationOfTorusUnstableFluxRopes} as an aid to interpret the observed evolution of the AR during the eruption. In their MHD model, the launch of the flux rope is the result of a loss of equilibrium when the threshold of the torus instability has been reached~\citep{KliemToeroek:TorusInstability}.

In the present work, the persistent twisting of the magnetic arcade in the core of the AR leads to the ejection of a series of flux ropes. Similarly recurrent plasmoid/flux rope ejections were modeled by~\citet{Manchester:EruptionOfEmergingFluxRope}, who carried out compressible MHD simulations of the emergence of a twisted magnetic flux tube from an idealized convection zone into a non-magnetized corona. The simulations showed that shear flows at the photospheric level driven by the dynamics of twisted emerging flux lead to recurrent flux rope ejections. In this sense, the present MF simulations provide a similar qualitative result.

Another similarity between the present model and MHD models of flux rope ejection is the presence of a X-point type topological feature between the ejecting flux rope and the underlying sheared arcade. Such a feature can be seen near the location $(y,z) = (0,20)$ Mm in the left panel of Fig.~\ref{fig:fig4}. The combined structure consisting of the sheared arcade, X-point and flux rope were also reported by~\citet{Savcheva:2012}, who used both an MHD model and a MF model to study the pre-eruptive magnetic field configuration of an AR. In their work, the formation of such a structure in the MHD model was the result of imposing converging flows as well as flux cancellation at the polarity inversion line. In comparison, the formation of similar structures in the corresponding MF model was the result of the relaxation of a current-carrying flux rope inserted into an initially potential field configuration. 

\subsection{Increase of horizontal photospheric field near the polarity inversion line following flux rope ejection}
\begin{figure*}
\centering
\includegraphics[width=\textwidth]{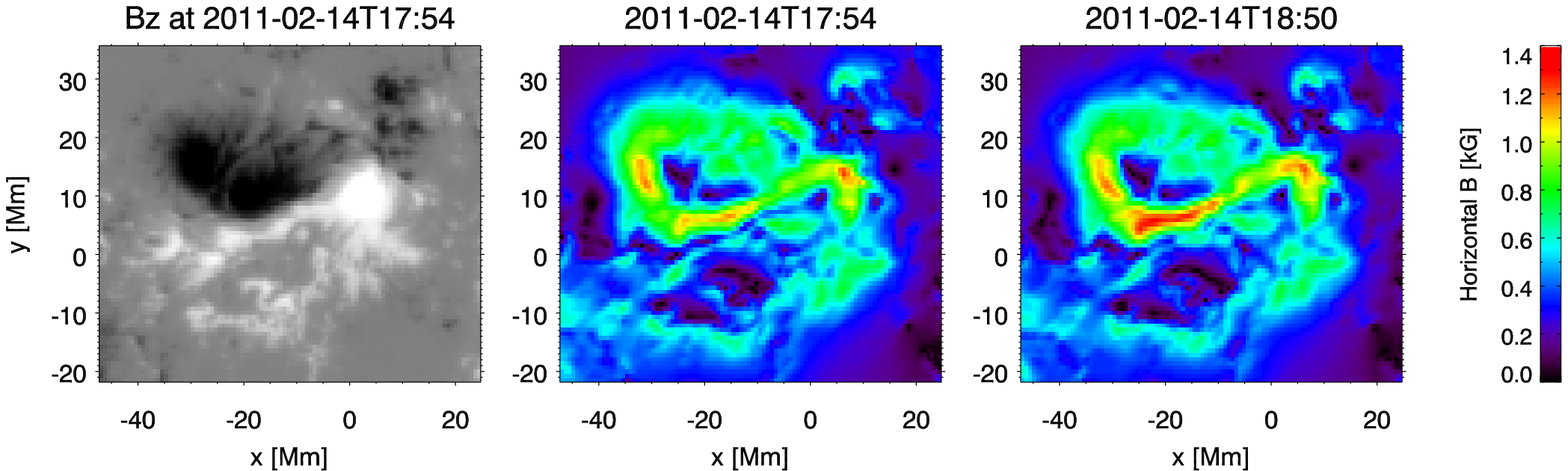}
\caption{Evolution of the horizontal component of the magnetic field in the core of the modeled AR 11158 (simulation run with $\Omega = 1/4$ turns per day). The field is sampled from the midplane of the lowest grid layer in the simulation at $z=364$ km. The ejection of a flux rope between 2011-02-14T17:55 and 2011-02-14T18:50 results in an increase of the horizontal field strength near the polarity inversion line.}\label{fig:collapse_plot}
\end{figure*}

An analysis of the vector magnetograms of AR 11158 taken by SDO/HMI shows a substantial increase in the strength of the horizontal field $B_h$ near the polarity inversion line after the X2.2 flare on Feb 15th 2011~\citep{Wang:ResponseOfPhotosphereAfterXFlare}. In a series of NLFFF extrapolations based on HMI vector magnetograms,~\citet{Sun:AR11158} found a downward displacement of the current-carrying channel above the polarity inversion line following the flare.

As indicated by Fig.~\ref{fig:collapse_plot}, a similar qualitative behavior can be found in simulation run AR11158$\Omega+0.25$ after the ejection of a flux rope. In the simulation, this increase in the horizontal field strength results from the relaxation of the arcade field following the magnetic reconnection event which produced the flux rope (see also animated version of Fig.~\ref{fig:fig4}, available online).

This episode of flux rope formation and arcade relaxation in the MF model occurs almost 8 hours before the actual X2.2 flare occurred in AR 11158 so the former should not be taken as a faithful representation of the observed flare and eruption. The discrepancy in timing between the two is not surprising given the use of only line-of-sight magnetograms and the imposition of an ad hoc uniform twisting rate in the model. However, there is one small benefit to not having used the full HMI vector magnetograms to drive the simulation. This omission rules out the possibility that the increase of $B_h$ in the model is simply a side effect of the increase of $B_h$ in HMI vector magnetograms. Instead, we can unequivocally attribute the $B_h$ enhancement in the simulation to the relaxation of the post-eruption arcade. The same physical mechanism, namely photospheric response to relaxation of a post-eruption arcade~\citep[mediated by the Lorentz force, see][]{Hudson:LorentzForce,Fisher:LorentzForce}, has been invoked by~\citet{Wang:ResponseOfPhotosphereAfterXFlare} to explained the observed increase of $B_h$ at the polarity inversion line in AR 11158. 

\section{Discussion}

In this paper, we present a framework for performing data-driven simulations of solar AR formation. Under the magnetofriction assumption, fluid velocities are assumed to be proportional to the local Lorentz force. This has the advantage that the velocity distribution is guaranteed to evolve the field toward a force-free state. Since the magnetic induction equation is solved to advance the magnetic field in time, \emph{ideal} magnetofrictional relaxation preserves the topology of the magnetic field while decreasing magnetic energy. In regions of high current density where resistive diffusion (used in this Eulerian code to facilitate magnetic reconnection) is important, the magnetic topology is allowed to change via reconnection. The change in the topology may permit the magnetic configuration to further relax to lower energy states under quasi-ideal, magnetofrictional evolution. 

By incorporating time sequences of photospheric magnetograms as boundary data, the method models changes in the coronal magnetic field in response to photospheric driving (including shearing flows and flux emergence). As an example of the application of this method to magnetogram data obtained by SDO/HMI, we performed a number of simulations to model NOAA AR 11158. Since the accurate retrieval of photospheric electric fields from temporal sequences of vector magnetograms is still a research problem~\citep{Fisher:EstimatingElectricFields}, we choose to perform the simulations with varying assumptions about the underlying $\vec{E}$ that is responsible for the magnetogram evolution. In the case when twisting motion is absent, flux rope ejections were produced by the model. When continuous twisting was imposed, the sheared field above the sharp polarity inversion line in the core of the AR produced a series of flux rope ejections. 

Since idealized assumptions were made about the nature of the underlying photospheric electric field, the simulations presented in this paper are not meant to be faithful representations of actual eruptions from  AR 11158. Nevertheless, this work demonstrates the potential utility of such a data-driven approach for modeling observed ARs. In a follow up paper, we will drive the model with actual electric fields retrieved from HMI vector magnetograms and constrained by observed Doppler velocities~\citep{Fisher:EfieldsWithDoppler}. The use of electric fields more faithful to the data will facilitate a side-by-side comparison of the modeled AR with EUV observations from AIA~\citep{Lemen:AIA}.

A method for synthesis of mock `coronal' images base on a proxy emissivity was introduced. The emissivity of points along a magnetic field line is assumed to proportional to the field-line average of $j^2$. This simple technique seems to produce with a visual texture similar to EUV images of coronal loops. While this technique is useful for visualizing an ensemble of coronal loops in a simple magnetic model, it is by no means a replacement for more sophisticated techniques that use the thermodynamic variables from MHD models and take into account atomic physics for EUV image synthesis.

\acknowledgements

M.C.M.C. and M.L.D. acknowledge support by NASA grant NNX08BB02G to LMSAL. Additionally, we acknowledge support by the Lockheed Martin SDO/HMI sub-contract 25284100-26967 from Stanford University (through Stanford University prime contract NAS5-02139). HMI is an instrument onboard SDO, a mission for NASA's Living With a Star program. This work was also made possible by NASA's High-End Computing Program. The simulation presented in this paper was carried out on the Pleiades cluster at the Ames Research Center. We thank the Advanced Supercomputing Division staff for their technical support.

\bibliographystyle{apj}
\bibliography{references,apj-jour}

\end{document}